# SECURE MULTIPATH ROUTING SCHEME USING KEY PRE-DISTRIBUTION IN WIRELESS SENSOR NETWORKS

Kamal Kumar[1], A. K. Verma[2] and R. B. Patel[3]

[1]M.M. Engineering College, Haryana, 133207 - India
[2]Thapar University, Patiala, Punjab, 147004- India
[3]G. B. Pant College of Engineering. Pauri, Uttrakhand, 131039- India

## *ABSTRACT*

*Multipath routing in WSN has been a long wish in security scenario where nodes on next-hop may be targeted to compromise. Many proposals of Multipath routing has been proposed in ADHOC Networks but under constrained from keying environment most seems ignorant. In WSN where crucial data is reported by nodes in deployment area to their securely located Sink, route security has to be guaranteed. Under dynamic load and selective attacks, availability of multiple secure paths is a boon and increases the attacker efforts by many folds. We propose to build a subset of neighbors as our front towards destination node. We also identified forwarders for query by base station. The front is optimally calculated to maintain the security credential and avail multiple paths. According to our knowledge ours is a novel secure multipath routing protocol for WSN. We established effectiveness of our proposal with mathematical analysis.*

## 1. INTRODUCTION

WSN network evolved as monitoring tool in adverse, dynamically changing environment. Besides being used for security critical applications it surfaced into daily life monitoring systems ranging from mining applications to steel furnace reporting, from libraries to under-water monitoring, from moisture controlling to dam-reservoir controller, from toll-plaza to grocery-stores. Applications are in abundance and so are the issues. With increasing demands for customized setups of WSNs new issues also surfaced. It needs custom solutions to demanding situations. Many works has addressed the specific issues in WSN. We specifically limit ourselves to problem of maintaining multiple routes not necessarily node-disjoint through networks which are equally secure and have qualified under a complex qualifying criterion in threat prone deployment areas. Besides improving upon best-effort delivery we tried to maintain high value of protection keys in the links. We are working on the principle that route is as strong as the weakest link in the route. We have proposed a probabilistic model for selecting our front towards a specific destination. Proposal has been generic and we specialized it to achieve security requirements in the threat prone deployments. Probabilistic model can be specialized for other requirements like distance, energy, throughput and delay. Section 2 discusses related work in the area with section 3 and 4presenting network and probabilistic analytical model of proposal and routing scheme. Section 5 we present performance analysis. With Section 6 we finally conclude and cite future directions.

## 2. RELATED WORKS

Several works in security and key management in WSN have been reported and most addressed the security of single path from sender to destination. Proposals in [2] and [3] addressed the the provisioning of security in WSN using Single Network-Wide Key. Compromising one key will compromise the security of whole network. In [3] an approach for establishing Pair-wise between every pair of nodes was proposed but the initialization of scheme is based on master key. Master key is erased after initialization and thus not scalable. A proposal in [55] proposed Trusted Base Station based key management scheme SPINS using SNEP and µTESLA as building blocks. SNEP offers data confidentiality, authentication, integrity, and freshness, while µTESLA offers broadcast data authentication. SPINS uses less of a sensor node's memory and the communication costs for SPINS are small, with security properties like data freshness, authentication and confidentiality. Several proposals [4] [5] [6] [8] addressed the security using key pre-distribution schemes. Most of these schemes allow the probabilistic approach to decide the security credential. In [7] and [9] authors proposed a scheme using deployment knowledge. A scheme in [10] was proposed for using location dependence in clustered hierarchical sensor networks. We have proposed few key management schemes in [11] [12] [13] [14] and [15]. Scheme in [11] guarantee connectivity using location effect in pre distributed keying environment. In [12] we proposed a scheme implementing framework for key management schemes in heterogeneous wireless sensor networks. In [13] we proposed a key management which is most computation efficient and storage efficient. In [14] we proposed a key management scheme which build secure route from source to destination using variance of keys on links on route while selecting next link on route being built. In [15] we implemented a key management scheme which exploits location information. None of the scheme cited above offers multiple and equally secure paths between source and destination.

## 3. OUR PROPOSAL

In this section we present our proposal with network elements and network model. We could address the query and data routing in our proposal using Query Relays (QR) and

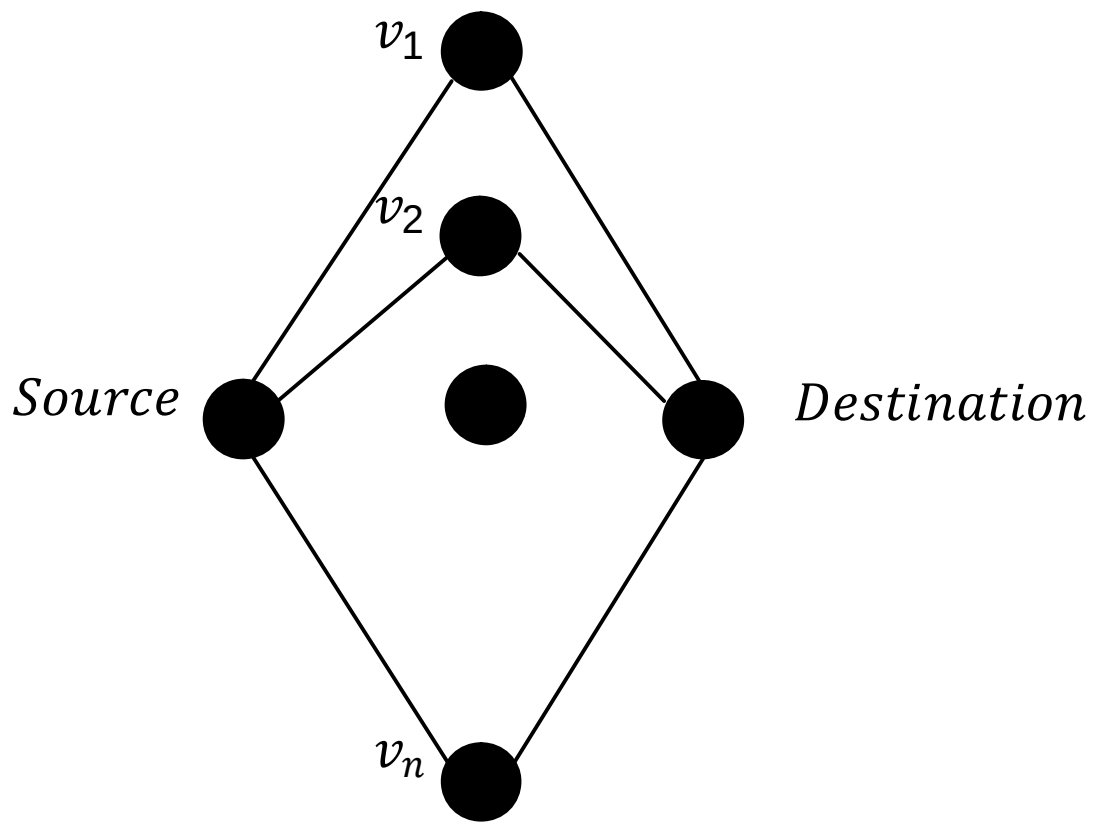

Figure 1. Wireless Environment Scenario

Data Relays DR). QR relays the query from sink to a deployment area or single node. DR route reply back to sink using data relays. The routes for query and reply may have same or disjoint routes and ensures minimum delay.

### 3.1 Preliminary

We consider a list of forward nodes as proposed in [1] for selecting nodes out of one hop neighbours towards a particular destination. This is as shown in figure.1. Single destination in WSN happens to be Base Station. Nodes are homogeneous in nature and have fixed transmission range. Diagram in figure represents an example scenario. Each link in figure1 cost some energy to sender and receiver. With error prone environment each link suffers some error. Node $u$ has selected nodes $\{v_1, v_2, \dots, v_n\}$ as possible set of forwarding nodes. This is treated as priority list and node $v_1$considered to most preferred node. Opportunistically nodes forward message sent by $u$ towards BS. There is possibility of multiple copies of message being forwarded by forwarder nodes because of hidden node problems. Opportunistic routing may suffer from duplicated packets as there is no solution for schedule for nodes forwarding packets via forwarder nodes and security is not considered and thus prone threats.

### 3.2 Network Model and Elements

We consider a wireless Sensor Network consisting of Large Number of L-Sensors and a few number of H-Sensors. Initially, we assume 15% of total nodes consist of H-Sensors. Nodes are given unique IDs and are assigned by Sink. Each Sensor has fixed transmission range. We may assume that H-Sensor has comparatively larger transmission range and more storage capacity to entail larger number of keys. The resultant network is modeled as multi-hop network and fits the definition of graph. An edge between a pair $SNs$ implies connectivity between concerned nodes.

Consider a WSN with nodes having unique identities (IDs). We assume that every wireless node $u$ has fixed transmission power. Assume $WSN = \{A, H, L, K, k_1, k_2, E, V\}$ where $A$ denotes deployment area dimensions, $H$ denotes number of H-Sensors, $L$ describes the strength of L-Sensors, $K$ is the key pool, $k_1$denotes number of keys given to L-Sensors, $k_2$denotes number of keys pre-distributed to H-Sensors. $E$ Denote the undirected/undirected edge set and $V$ denotes nodes set respectively. Each directed link $u \longrightarrow v$ has a nonnegative weight, denoted by $k(u, v)$ which is the number of shared randomly pre-distributed keys and to be used by node $u$ together to send a packet to node $v$ for encryption during forwarding. In addition, each link has a failure probability, denoted by $f(u, v)$, which is the probability that a transmission over link $(u, v)$ is not successful because of unavailability or schedule, i.e., to have a chance of $1 - f(u, v)$ for successful secure transmission a packet to node $v$; node $v$ must be active or not simultaneously receiving other transmission. No transmission is possible if node's shares no key. To illustrate the idea let us consider a network example in Figure 2.

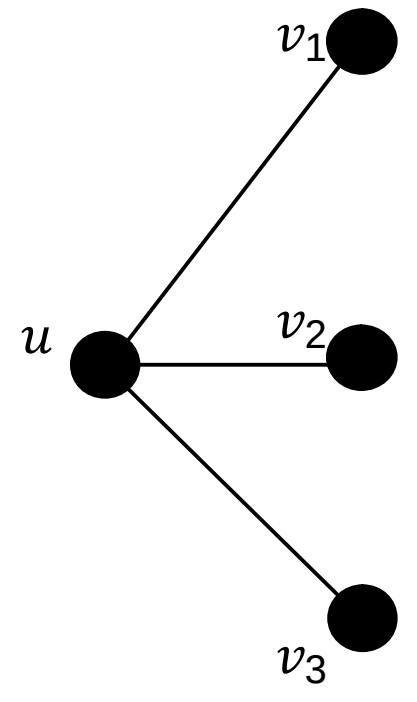

Figure 2. Example Scenario

The unavailability probability from the source node to each node $v_i$ is $f$ and is same for all links. In our proposal instead of relaying through one node, say $v_i$; we propose to use a set of nodes forming a forwarding relays which is a priority list for routing packets towards a fixed destination i.e. base station. We call such nodes forward relays ($fR$). We can compute that the expected number of transmissions will be $\frac{1}{1-f}$ for the intended node $v_i$ to receive the packet correctly.

Let node $v_i$ is selected as member of $fR$ by $v_j$, $v_k$ and $v_l$ nodes, such set of nodes i.e. $\{v_j, v_k, v_l\}$ is Selectors Set and we will use them as query relays and named as backward relays( $bRs$) on reverse path from sink to node(s).

On the other hand, by having multiple $fR$ to counter for unavailability or outage and considering multiple one-hop nodes in the role of $fR$, in symmetric-paired-key environment, the expected numbers of transmissions for at least one node to receive the packet by $fRs$ increase to $\frac{1}{(1-f)^n}$. The denominator term is raised to power $n$, because of paired keying environment in random key pre-distribution which contrasts from broadcast environment. Assume that $fR$ are maintained as priority list. The $fR$ list is prioritized to indicate which nodes have higher priority to forward the packet. The node in $fR$ list, which received the packet successfully, will act as new source nodes and route the packet to the target node via its $fR$. Finally; main idea of our secure forwarding which we named as Expected Secure Relaying (ESR) is as follows: we let $EAK_u(fR(u))$ denote the Effective Average Key harnessed on the route from node $u$ to $Sink$, where $fR(u)$ is chosen by $u$ as $fR$. During initial step $EAK_{Sink}$ is initialized to 0 along with all other nodes. The updates on $EAK_u$, $fR(u)$ and $bR(u)$ are computed periodically.

### 3.3 Setting up Forward Relay Key ( $fR_{key}$ )

Equation $\frac{1}{(1-f)^n}$ specifies the number of retransmissions to be performed for at least one in $fRs$ receive and forward the packet from its selector. If we can increase the denominator to $(1-f^n)$ by providing a broadcast environment with nodes in $fR$ we can reduce number of broadcast for at least one node in $fR$ receive successfully. In case of encryption using all pair-wise keys obtained it is difficult to have broadcast environment. We strive to establish a secure broadcast key between node and its $fR$ using following method. Let we identify nodes $i, j$ as nodes in $fR$ of node $s$. Figure 3 and 4 shows the steps for establishing forward key. Consider node $s$ being selector sends encrypted messages to members $i$ & $j$ in steps 1 to 6 are performed in sequence as shown in figure above.

Messages in step 1 and 2 are encrypted using all pre-distributed shared keys between $i-s$ and $j-s$ pairs. On verifying the integrity of messages; $i$ & $j$ compute their shares individually and sends messages to $s$ encrypted using all pre-distributed shared keys between $i-s$ and $j-s$ in steps 3 and 4 respectively. Having received all shares from $fR(s)$ selector node $s$ generates its unique share and using X-OR of all shares with its own share generates a unique $fR_{key}$ for communication with $DRs$ only. In step 5 and 6 selector node $s$ dispatches $fR_{key}$ key message destined for $i$ & $j$. Now $i$ &$j$ can generate $fR_{key}$ key using the contents from $fR\ Key$ message and own share there by verify the identity of sender and integrity of message. Following equation (1) gives an insight of operation:

$$fR_{key} = SH_s \oplus SH_i \oplus SH_j \qquad (1)$$

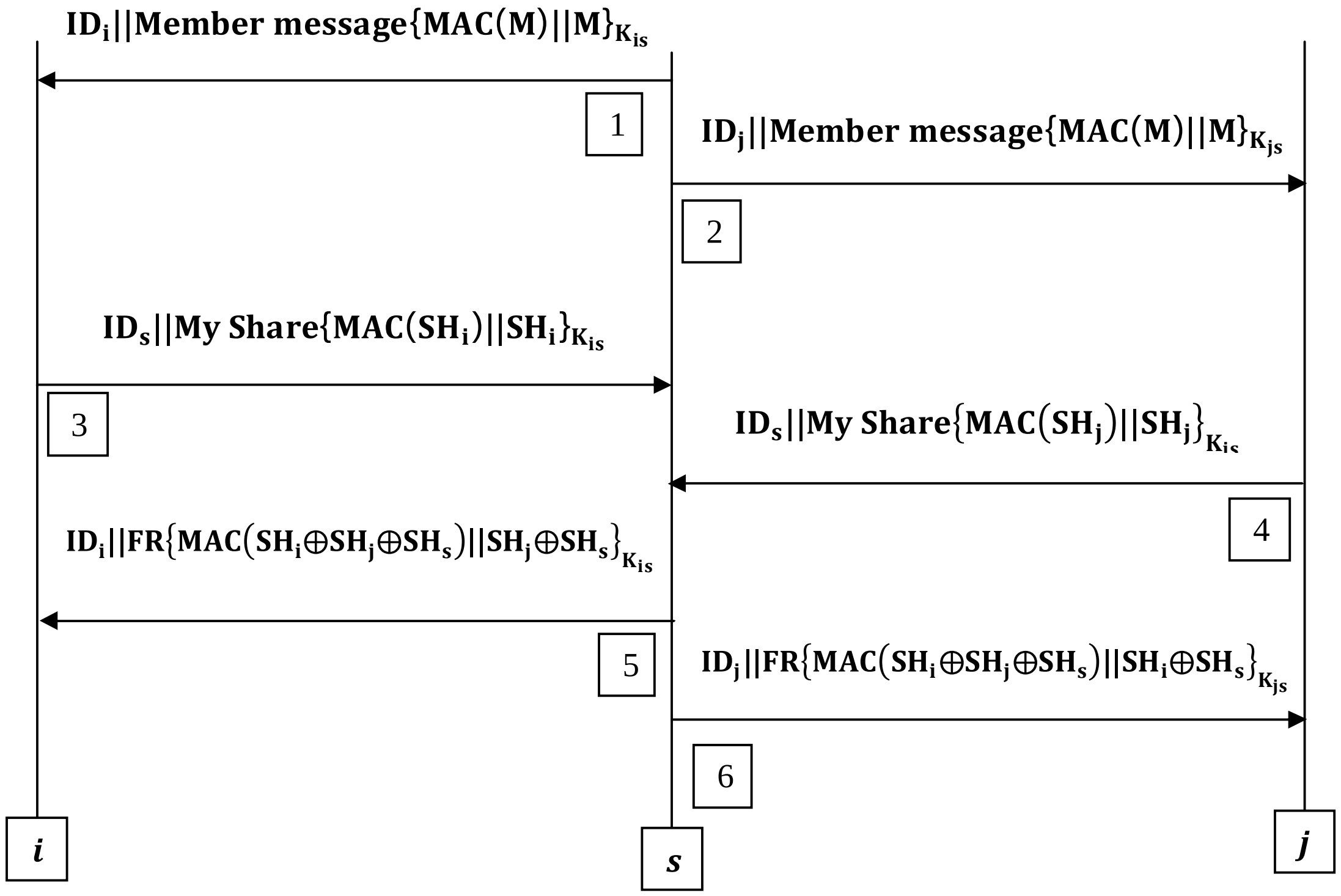

Figure 3. Forward Key Establishment from Shares

**Step 1**: $ID_i||Member\ message\{MAC(M)||M\}_{K_{is}}$
**Step 2**: $ID_j||Member\ message\{MAC(M)||M\}_{K_{js}}$
**Step 3**: $ID_s||My\ Share\{MAC(SH_i)||SH_i\}_{K_{is}}$
**Step 4**: $ID_s||My\ Share\{MAC(SH_j)||SH_j\}_{K_{js}}$
**Step 5**: $ID_i||\ FR\{MAC(SH_i \oplus SH_j \oplus SH_s)||SH_j \oplus SH_s\}_{K_{is}}$
**Step 6**: $ID_j||FR\{MAC(SH_i \oplus SH_j \oplus SH_s)||SH_i \oplus SH_s\}_{K_{js}}$
**Step 7**: $fR_{key} = SH_s \oplus SH_i \oplus SH_j$

Figure 4. Steps for Establishing Forward Key

Thus $fR_{key}$ key is established using shares from contributors. As the numbers of forwarders are one or more so are the contributions. To compromise the $fR_{key}$ every path between selector and it's $fRs$ has to be compromised. As $fR_{key}$ is common among selectors and $fRs$; we now are able to exploit the broadcast advantage in wireless medium and reduce the number of trials for at least one of forwarders receive and forward the packet. Increasing the denominator in equation from $(1-f)^n$ to $1-f^n$ will decrease the number of trials for successful receiving and forwarding of message.

### 3.4 Setting up Backward Relay Key $bR_{key}$

Consider that a particular node has obtained distinct $fR_{key}$ for use with distinct selector. Assume that a node $i$ was in $fRs$ of set $S$ where $S = \{s_1, s_2, s_3\}$. We identify set $\{s_1, s_2, s_3\}$ as possible backward Relay ($bR$) set of node $i$. Let $bR(i)$ denotes $bR$ set of node $i$ and $bR(i) = \{s_1, s_2, s_3\}$. Assuming $fR_{key}^{s_1,i}$ , $fR_{key}^{s_2,i}$ and $fR_{key}^{s_3,i}$ denotes $fR_{key}$ between $s_1 - i$, $s_2 - i$ and $s_3 - i$ , node $i$ can now compute $bR_{key}$ for use as query broadcast key by node $i$. Following steps outline the generation and distribution of $bR_{key}$.

Step 1: $ID_i || \; bR\; SetUp\{MAC(Q)||Q\}_{fR_{key}^{s_1,i}}$

Step 2: $ID_i || \; bR\; SetUp\{MAC(Q)||Q\}_{fR_{key}^{s_2,i}}$

Step 3: $ID_i || \; bR\; SetUp\{MAC(Q)||Q\}_{fR_{key}^{s_3,i}}$

Step 4: $ID_i || \; Shares\{MAC(fR_{key}^{s_2,i} \oplus fR_{key}^{s_3,i}) || fR_{key}^{s_2,i} \oplus fR_{key}^{s_3,i}\}_{fR_{key}^{s_1,i}}$

Step 5: $ID_i || \; Shares\{MAC(fR_{key}^{s_1,i} \oplus fR_{key}^{s_3,i}) || fR_{key}^{s_1,i} \oplus fR_{key}^{s_3,i}\}_{fR_{key}^{s_2,i}}$

Step 6: $ID_i || \; Shares\{MAC(fR_{key}^{s_1,i} \oplus fR_{key}^{s_2,i}) || fR_{key}^{s_1,i} \oplus fR_{key}^{s_2,i}\}_{fR_{key}^{s_3,i}}$

Step 7: $bR_{key}^{bR(i),i} = (fR_{key}^{s_1,i} \oplus fR_{key}^{s_2,i} \oplus fR_{key}^{s_3,i})$

Figure 5. Steps for Establishing Backward Key

Figure 5 shows the steps for establishing backward key. Step 1 to step 3 is encrypted communication from node $i$ to each of it's $bR(i)$ for $bR_{key}$ key setup. Step 4 to step 6 results into dispatch of partial key to selectors. Step 7 finally establishes $bR_{key}$ at node $i$ and selectors $\{s_1, s_2, s_3\}$.

### 3.5 Expected Key Average

Now, we present the main idea of calculating the Effective Key Average ($EKA$) for each node and selecting the forward Relays $(fR)$. We define $EKA$ as the average keys used to provide a broadcast environment in pre-distributed keying environment. As in above section $fR_{key}$ has been established using all shared pre-distributed keys on links between selectors and $fR$ nodes, which implies that effectiveness of routing in our customized broadcast environment using $fR_{key}$ is as effective as is the average number of keys used in setting up of $fR_{key}$.

Consider a node $u$ and its one-hop neighbors. We will compute the, $EKA$ and $NHList(u)$ of node $u$ based on the $EKA$ of its neighbors whose $EKA$ of sending data to the $Sink$ has already been computed. We want to choose a subset of neighboring nodes $N(u)$ as $NHList(u)$ of node $u$ such that the $EKA$ on the route from node $u$ to send a packet to $Sink$ is maximized.

Consider $Sink$ as our destination node. Given a set of nodes $U$, let $U^{\#}$ defines the sorted list of $U$ based on $EKA$ to send data (via possible relay) to $Sink$.

If $NHList(u)$ denote the priority next hop list of node $u$ then $NHList^{\#}(u)$ represents sorted next hop list on $EKA$ in decreasing order. i.e, $NHList^{\#}(u) = \{v_1, v_2, \dots, v_{|NHList(u)|}\}$ where $i < j \Longrightarrow EKA_{v_i} \leq EKA_{v_j}$ .Using the theory of probability let $F$ denotes the probability of total failure i.e. a packet sent by node $u$ is not received by any node in $NHList^{\#}(u)$. Clearly,

$$F = \prod_{i=1}^{|NHList^{\#}(u)|} f_{uv_i} \quad (2)$$

The probability of at least one node in $NHList^{\#}(u)$ will receive packets successfully, can be computed as $R = 1 - F$. We can compute the number of trials that node $u$ must perform in order to achieve first success by $1/_R$ . For e.g. if probability of success is 0.5 then number of trial to have first success can be given by $1/0.5 = 2$. If $1/_R$ gives the number of trials that a node must perform to send a packet which is received by at least one in the $NHList^{\#}(u)$ then using trials information for nodes $NHList^{\#}(u)$ we can compute possible delay incurred to get the packet at $Sink$.

Let, $EKA_u^{v_i}(NHList^{\#})$ denote the expected key average on next hop from $u$ through one of the node $v_i$in $NHList^{\#}$ then $EKA_u^{v_i}$ that will be used can be computed as:

$$EKA_u^{v_i}\left(v_i \in NHList^{\#}(u)\right) = \frac{k_{uv_1} * \left(p_{uv_1}\right) + k_{uv_2} * \left(p_{uv_2}\right) + \cdots + k_{uv_{|NHList^{\#}(u)|}} * \left(p_{uv_{|NHList^{\#}(u)|}}\right)}{1 - f^{|NHList^{\#}(u)|}} \quad (3)$$

Where $p_{uv_i}$represent the probability of forwarding to $v_i$ in$NHList^{\#}(u)$. As one of the forwarder has to forward ultimately (may require many trials) requires that $\sum_{i=1}^{|NHList^{\#}(u)|} p_{uv_i} = 1$. For e.g. If $|NHList^{\#}(u)| = 3$ then $NHList^{\#}(u) = \{v_1, v_2, v_3\}$. If forwarder $v_1$has been assigned highest priority of among three forwarders with $v_2$assigned second then we have $4X + 2X + 1X = 1$ where $p_{uv_i} = 2 * (i - 1) * X$ . This implies $p_{uv_1} = 4 * X, p_{uv_2} = 2 * X$ , $p_{uv_3} = 1 * X$. This leads to $p_{uv_1} = 0.57, p_{uv_2} = .29,\ p_{uv_3} = .14$. If $f(u, v_i) = .5$
If we assume that $k_{uv_1} = 30, k_{uv_2} = 27,\ k_{uv_3} = 22$ then $C_u^{v_i}(NHList^{*}) = 54$.

When at least one node in the forwarder list of node $u$ received the packet successfully, we need to calculate the expected cost to forward the packet sent by node $u$. Let $EKA_u^{Sink}\left(NHList^{\#}(u)\right)$ denotes $EKA$for $u$ to forward (using some nodes in the forwarder list of $u$) the packet to the $Sink$.If $EKA_u^{Sink}\left(NHList^{\#}(u)\right)$ represent Effective Average Keys of route through $NHList^{\#}(u)$ can be calculated as follows: assume the relays list is$NHList^{\#} = \{v_1, v_2, \dots, v_{|NHList(u)|}\}$. The probability that node $v_1$ forwards the packet is $1 - f(u, v_1)$ and Effective Keys Average of $v_1$is $EAK_{v_1}^{Sink}$; then node $v_2$ will forward the packet with probability $f(u, v_1) * (1 - f(u, v_2)$ and the Effective Keys Average will be $EAK_{v_2}^{Sink}$ . Basically, node $v_i$ forwards the packet if it receives the packet and nodes $v_j$; $0 < j < i$ did not receive the packet, and in this case, the Effective Average Keys will be $EAK_{v_i}^{Sink}$. Hence, $EKA_{v_i}^{Sink}$ can be computed as follows:

$$EAK_u^{Sink}\left(v_i \in NHList^{\#}(u)\right) = \left(1 - f_{uv_1}\right) * EAK_{v_1}^{Sink} + \sum_{i=2}^{|NHList^{\#}|}\left(\prod_{j=1}^{i-1} f_{uv_j}\right) * \left(1 - f_{uv_i}\right) * EAK_{v_i}^{Sink} \quad (4)$$

Finally; $EAK_u(NHList^{\#})$ the on route from $u$ to $Sink$ is computed as follows:

$$EAK_u\left(NHList^{\#}(u)\right) = \frac{EAK_{v_i}^{Sink}\left(v_i \in NHList^{\#}(u)\right)}{1 - f^{|NHList^{\#}(u)|}} \quad (4.1)$$

$$EAK_u^{Sink} = EAK_u^{v_i}\left(v_i \in NHList^{\#}(u)\right) + EAK_u\left(NHList^{\#}(u)\right) \quad (4.2)$$

Equation (4) illustrated how to compute $EAK$ of a sender to broadcast a packet if the current chosen forwarder list is $NHList^{\#}(u)$. Equation (4.1) computes tentative $EAK$ which finalizes $NHList^{\#}(u)$ and equation (4.2) computes real $EAK$ by augmenting tentative $EAK$ with last-hop cost computed in equation (3). Thus first part i.e. equation (4.1) is $EAK$ for the sender to successfully transmit a packet to at least one receiver in $NHList^{\#}$. The second part i.e. (4.2) corresponds to $EAK$ that one of node in the $NHList^{\#}$ finally to relays the packet to the final destination node.

### 3.6 Finding the $NHList$

Instead of random selection of nodes from $N(u)$, we choose a prefix of sorted neighbor list $N^{\#}(u)$ as our result i.e. $NHList^{\#}(u)$ . For a given $N^{\#}(u)$ there can be at the most $|N^{\#}(u)| + 1$ prefixes. Selecting nodes from $N^{\#}(u)$ , one at each step provided $EAK_{v_i} > EAK_u$ . If $v_i$ fails to satisfy the required condition; every node ahead of $v_i$ in $N^{\#}(u)$fails to satisfy the said condition.

## 4. ROUTING ALGORITHM

How nodes will select their forwarder list and how to use expected cost is highlighted in previous section. Now we are able standardize the steps as collection of three algorithms, namely; $Update_EAK$, $Compute_EAK_to_Sink$, $Dispatch_NHList$. These algorithms are presumed to be hardcoded and can be executed as per their requirements. After execution of $Compute_Cost_to_Sink$ sink has information of about selectors and Relays. Each selector may have multiple relays and each node may possess multiple selectors. In each case we have a subset of one-hop neighbors as selector or relays or selectors-relays combined. Using the information received from nodes in deployment area sink is able to compute routes from sink to nodes. Sink may use these routes to periodically diffuse query in the network, whereas nodes may use their forwarders towards sink to report any urgent event. The algorithm's pseudo code is described in figure 6.

## 4.2 Exchanging ($NHList$ ) List Information

Each node prioritized their relays in $NHList$. Selection along with priority is informed to relays by selectors. This process may be initiated by nodes after completing the execution of

$Compute_EAK_to_Sink\left(Sink, V, EAK_u(NHList^{\#}(u))\right)$

$BEGIN\ \{Compute_EAK_to_Sink\}$

$Sink_Broadcast_Initialize$

$Node_Initialize: EAK_u(NHList^{\#}(u)) = 0, NHList^{\#}(u) = \phi$

$Node_Broadcast_IDS + EAK$

$Node\ Sort\ Neighbour\ List\ on\ EAK$ in decreasing order to get $N^{\#}$

$Nodes\ Executes: EAK_u(NHList^{\#}(u)) = Update_EAK\left(EAK_u(NHList^{\#}(u)), N(u)\right)$

$u \in V, \quad EAK_u(NHList^{\#}(u)) = 0, \quad EAK_{Sink}^{Sink}(NHList^{\#}(Sink)) = 0$

$Sink_Limited_Broadcast\{EAK_{Sink}^{Sink}\}$

$\forall u \in N(Sink) Executes:\ EAK_u^{Sink}(NHList^{\#}(u)) = Update_EAK\left(EAK_u^{Sink}(NHList^{\#}(u)), N(u)\right)$

$repeat$

$let\ S_1 = V - \{Sink\}\ and\ S_2 = \{Sink\}$

$repeat$

$v = min_cost\{S_1\}$

$S_1 = S_1 \cup \{v\}\ and\ S_2 = S_2 - \{v\}$

$\forall u \in N(v) \cap S_1 : EAK_{TEMP} = EAK_u^{Sink}(NHList^{\#}(u))$

$\forall u \in N(v) \cap S_1 : NHList^{\#}(u)_{TEMP} = NHList^{\#}(u)$

$\forall u \in N(v) \cap S_1\ Executes: EAK_u(NHList^{\#}(u))$

$= \quad Update_EAK\left(EAK_u(NHList^{\#}(u)), N(u)\right)$

$until\ \ S_1 = \emptyset$

$\forall u \in V, Node_Broadcast_to_N(u): EAK_u^{Sink}(NHList^{\#}(u))$

$\forall u \in V\ Executes: EAK_u(NHList^{\#}(u))$

$= \quad Update_EAK\left(EAK_u(NHList^{\#}(u)), N(u)\right)$

$until\ \ No\ Change\ in\ EAK\ \ and\ NHList^{\#}$

$Dispatch_NHList^{\#};$

$End\ \{Compute_EAK_to_Sink\}$

---

$Update_EAK\left(EAK_u(NHList^{\#}(u)), N(u)\right)$

$BEGIN\ \{Update_EAK\}$

Sort the neighboring nodes $N(u) = \{v_1, v_2, \dots, v_{|N(u)|}\}$ based on their EAK in decreasing order and get $N^{\#}(u)$.

$for(i = 1; i < |N^{\#}(u)|; i++)$

$if\left(EAK_u(NHList^{\#}(u)) < EAK_{v_i}^{Sink}(v_i \notin NHList^{\#}(u))\right) then$

$NHList^{\#}(u) = NHList^{\#}(u) \cup \{v_i\}$ and update $p_{uv_i}$

Update $EAK_u(NHList^{\#}(u))$ using equation(5) in steps (5.1) and (5.2)

$return\left(EAK_u(NHList^{\#}(u))\right)$

$End\ \{Update_EAK\}$

---

$Dispatch_NHList^{\#}()$

$BEGIN\ \{Dispatch_NHList^{\#}\}$

$u \in V, u \Rightarrow NHList^{\#}(u)$

$End\{Dispatch_NHList^{\#}\}$

**Figure 6** Routing Scheme

$Compute_EAK_to_Sink$. Relays node in $NHList$ are like vectors disclosing direction towards sink. Reverse channel is always available. Now relays have information of their relays and selectors. This information is propagated to sink using unicast messages through relays in $NHList$. Aggregating the information by relays nodes help reduce the number of messages. Selectors are proposed to be used for routing any query towards a region or node and path through relays to route a reply to destination sink respectively. We have classified the Selector nodes as Query Forwarder and Relay Nodes as Data Forwarders.

### 4.3 Route Construction

Sink has information of node wise selectors and relays. For query forwarding sink constructs query route using pairs like:

$(D, \{Fwd(D)\})$,
$(\{Fwd(D)\}, \{Fwd(\{Fwd(D)\})\})$
…
$(Fwd(\{\ldots\{Fwd(\{Fwd(D)\})\}\}), Sink)$.

Each such pair gives a possible hop on the respective paths. As a result sink may obtain all possible paths towards a specific node i.e. D or vice-versa. $Sink$ may choose any of such paths for propagation of query. Sink may choose any of the route on the basis of optimization criterion which may be delay, energy, hop count or else. Query with specified route is encrypted /decrypted on the path as it travels from sink to D.

## 5. PERFORMANCE ANALYSIS

In this section we present a simple and effective validation of our schemes using theorems.

**Theorem 1:** $NHList^{\#}(u)$ of node u must be a prefix of $N^{*}(u)$.
Proof: we proof this theorem by contradiction. Let $v_k$, $v_{k+1}$ are two nodes such that node $v_{k+1}$is in $L^{\#} = L \cup \{v_{k+1}\}$ and $v_k$is not. Let $EAK_u(L^{\#})$is expected key value after and $EAK_u(L)$is expected key values before considering $v_{k+1}$. Let $\Delta_{k+1}$ represent the increment achieved, i.e. $EAK_u(L^{\#}) = EAK_u(L) + \Delta_{k+1}$. Had it been $v_{k+1}$ then $L^{\wedge} = L \cup \{v_k\}$. In $N^{\#}(u)$ , $v_k$ comes earlier than $v_{k+1}$as $N^{\#}(u)$is sorted on effective key averages. This implies $\Delta_k \geq \Delta_{k+1}$ and$EAK_u(L^{\wedge}) \geq EAK_u(L^{\#})$ . Thus selection of $v_{k+1}$ ahead of $v_k$ contradicts our selection criterion. Hence $NHList^{\#}(u)$ is prefix of $N^{\#}(u)$.

We further study the properties of forwarder list by introducing another three theorems. The first theorem, Theorem 2, shows that if a node, whose expected cost is less than the expected cost of a prefix forwarder list, is added to the forwarder list, then the expected cost of the newly created forwarder list will decrease while it will still be greater than the expected cost of the newly added node. The second theorem, Theorem 3, shows that if a node, whose expected cost is greater than the expected cost of a prefix forwarder list, is added to the forwarder list, then the expected cost of the newly created forwarder list will increase. Theorem 4 establishes connectivity issues.

**Theorem 2**: Consider a node $u$, a prefix $NHList^{\#}$ and a node $v_k \in N(u)/NHList^{\#}$. if $EAK_{v_k} > EAK_u(NHList^{\#})$ then $EAK_u(NHList^{\#} \cup \{v_k\}) > EAK_u(NHList^{\#})$ and $EAK_u(NHList^{\#})$ is monotonically non-decreasing.

Proof: We can prove above theorem by induction. Let us assume that node to be considered first from $N^{\#}(u)$, is $v_1$ and $L^{\#} = L \cup \{v_1\}$. Let us assume that $|L| = 0$ and $EAK_u(L) = 0$. Using equation (5.1) $EAK_u(L^{\#}) = EAK_u(L) + \Delta$ where $\Delta = (1-f) * EAK_{v_1}\left(L^{\#}(v_{k+1})\right)$ and $f$ represents non-negative error probability. This implies $EAK_u(L^{\#}) \geq EAK_u(L)$.

**Induction step:** Considering $v_k$ next from $N^{\#}(u)$, is $v_k$ and $L^{\#} = L \cup \{v_k\}$. Let us assume that $|L| = k-1$ and $EAK_u(L)$ is expected key average earned. $EAK_u(L^{\#}) = EAK_u(L) + \Delta$ Where $\Delta = f * f * \ldots (k-1\, terms) * (1-f) * EAK_{v_k}\left(L^{\#}(v_k)\right)$ and $f$ represents non-negative error probability. This implies $EAK_u(L^{\#}) \geq EAK_u(L)$. Hence, adding next node increments $EAK$ in case $v_k$ qualifies feasibility criterion of being a member in $NHList^{\#}$ .

Considering next from $N^{\#}(u)$, is $v_{k+1}$ provided $EAK_{v_{k+1}} \geq EAK_u$ and$L^{\#} = L \cup \{v_{k+1}\}$. Let us assume that $|L| = k$ and $EAK_u(L)$ is expected key average earned. $EAK_u(L^{\#}) = EAK_u(L) + \Delta$ where $\Delta = f * f * \ldots (k\, terms) * (1-f) * EAK_{v_{k+1}}\left(L^{\#}(v_{k+1})\right)$ and $e$represents non-negative error probability. This implies $EAK_u(L^{\#}) \geq EAK_u(L)$. Hence, adding next node increments $EAK$ in any case.

**Theorem 3**: Querying any node $u \in V$ will reach concerned $u$ in O (n) time.
Proof: As Sink has information about relays and selectors in the network. Sink computes all possible paths towards $u$. Sink unicast the query consisting of route to $u$ to node at one-hop. One hop nodes sends query to one of his selectors mentioned in the path. During query forwarding process relay nodes (selectively) forwards query to selector mentioned in the path of query. Query follows specified path in the network, and reaches $u$ in limited number of hops. As in the worst case path length is $(N-1)$. Reply node becomes new source of reply and will route reply on encrypted paths through its Data-Relays.

**Theorem 4:** All Nodes $(\forall u \in V)$ in the network are reachable.
Proof: Let we prove theorem by contradiction. Let there be a node $u$ which is unreachable as there is no route to $u$ at sink. This implies $u$ is not selector of any node. It implies$NHList(u) = \emptyset$. With no doubt we can be concluded that $N(u) = \emptyset$. This suggests a partitioned network. Otherwise; in a connected network $\forall u \in V$, $N(u) \neq \emptyset$ and equation (5) ensures that only neighbour will be in $NHList(u)$. Thus, in a connected network we have $NHList(u) \neq \emptyset$. As a fact sink will have route(s) to $\forall u \in V$.

## 6. Conclusion and Future Directions

We have proposed a new kind of multi path secure routing which distinguishes relays for query and reply, classified as Data-Relays (DRs) and Query-Relays (QRs). With provision of multiple DRs and QRs we have reduced the number of trials for successful traversal of packets from source to sink. The optimal selection of DRs in the network has been proposed, with objective of maximizing the Effective Average Keys on the routes from random node to sink. As the route was specified by sink and Forwarders are selected by nodes on the path, any masquerading and modification attack rendered ineffective. The analytical modelling supported the objectives and supports the strength of proposal. An implementation of the scheme is our next assignment. The scheme may be specialized for study of different parameters in demanding environments.

## AUTHORS

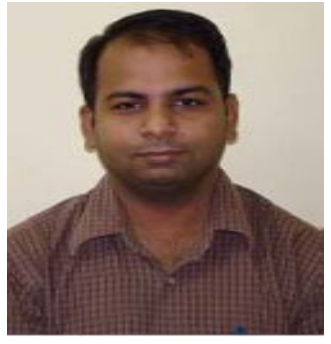

**Kamal Kumar** received his M.Tech. as well as B.Tech degree from Kurukshetra University, Kurukshetra, India.Presently he is working as Associate Professor in Computer Engineering Department in M.M. Engineering College, Ambala, India. He is pursuing Ph. D from Thapar University, Patiala, India.

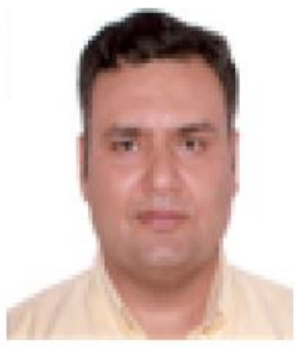

**A. K. Verma** is currently working as Associatet Professor in the department of Computer Science and Engineering at Thapar University, Patiala in Punjab (INDIA). He received his B.S. and M.S. in 1991 and 2001 respectively, majoring in Computer Science and Engineering. He has worked as Lecturer at M.M.M. Engg. College, Gorakhpur from 1991 to 1996. From 1996 he is associated with the same University. He has been a visiting faculty to many institutions. He has published over 80 papers in referred journals and conferences (India and Abroad). He is member of various program committees for different International/National Conferences and is on the review board of various journals. He is a senior member (ACM), LMCSI (Mumbai), GMAIMA (New Delhi). He is a certified software quality auditor by MoCIT, Govt. of India. His main areas of interests are: Programming Languages, Soft Computing, Bioinformatics and Computer Networks. His research interests include wireless networks, routing algorithms and securing ad hoc networks.

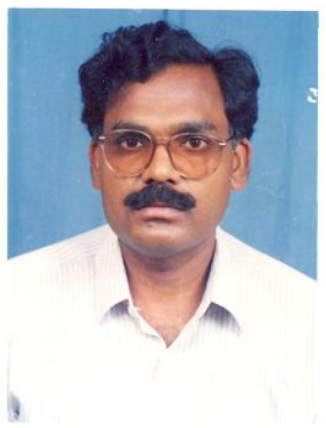

**R. B. Patel** received a PDF, Highest Institute of Education,Science & Technology (HIEST), Athens, Greece, 2005. He received a PhD in Computer Science and Technology from Indian Institute of Technology (IIT), Roorkee, India. He is member IEEE, ISTE. His current research interests are in Mobile and Distributed Computing, Security, Fault Tolerance Systems, Peer-to-Peer Computing, Cluster Computing and Sensor networks. He has published more than 100 papers in International Journals and Conferences and 17 papers in national journal/conferences. Two patents are also in the credits of Dr. Patel in the field of Mobile Agent Technology and Sensor Networks.